# MODELING NEEDS FOR HIGH POWER TARGET

**Charlotte Barbier[1*], Sujit Bidhar[2], Marco Calviani[3], Jeff Dooling[4], Jian Gao[5], Aaron Jacques[1], Wei Lu[1], Roberto Li Voti[6], Frederique Pellemoine[2], Justin Mach[1], David Senor[7], Fernando Sordo[8], Izabela Szlufarska[9], Joseph Tipton[1], Dan Wilcox[10], Drew Winder[1]**

[1] Oak Ridge National Laboratory, Oak Ridge, Tennessee, USA
[2] Fermi National Accelerator Laboratory, Batavia, Illinois, USA
[3] European Laboratory for Particle Physics (CERN), Geneva, Switzerland
[4] Argonne National Laboratory, Lemont, Illinois, USA
[5] Facility for Rare Isotope Beams, Michigan State University, Michigan, USA
[6] INFN, Sezione di Roma e Università degli Studi di Roma "Sapienza," Italy
[7] Pacific Northwest National Laboratory, Richland, Washington, USA
[8] ESS-Bilbao, Derio, Spain
[9] University of Wisconsin, Madison, Wisconsin, USA
[10] Rutherford Appleton Laboratory, Oxfordshire, UK

[*] *Corresponding author:* barbiercn@ornl.gov

1. **Introduction**

The most common target design is a solid material where the composition is generally defined by the physics requirements (in other words, what you want to produce). It is cooled by conductive or passive radiative cooling. However, this traditional design has already fallen short in the case of a higher average beam power because proper cooling is needed to operate safely. For instance, high power neutron target designs include flowing liquid metal [1] [2] [3], and the next generation designs are using rotating targets (European Spallation Source in Lund, Sweden, and the planned Second Target Station, in Oak Ridge, Tennessee, USA) to cool the target material properly. Although liquid metal targets have been successful overall, they lead to several operational challenges as a result of (1) unpredictable failure caused by cavitation erosion, weld failure, or high-cycle fatigue; (2) long fabrication time due to their complex geometry and weld requirements; and (3) the high cost associated with fabrication, storage, and disposal. Inert gas bubbles have been injected into the liquid metal to mitigate the pressure wave and induced-cavitation erosion. Nevertheless, this method will be appropriate only up to a specific power before cavitation erosion appears again. The upcoming rotating solid targets will have fabrication and cost issues similar to the liquid metal targets as well as the pitfalls associated with moving parts, but hopefully they will have a longer life span. Today the reliability of a high power target is already a challenge: major high power accelerator facilities (Oak Ridge National Laboratory's Spallation Neutron Source [SNS], Japan Proton Accelerator Research Complex's Materials and Life Science Experimental Facility, Fermi National Accelerator Laboratory's [FermiLab's] Main Injector Neutrino Oscillation Search [MINOS]) have had to operate at limited beam power because of target concerns. Thus, R&D associated with improving the reliability of high power targets is imperative and should not be underestimated.

Simulations play a critical role in target design (and the design of other accelerator components such as beam dumps, collimator, absorbers, scrapers): they predict both the beam-matter interactions and the thermo-mechanical response induced by beam heating, evaluate the radiation damage with time, and ensure safe operation. In general, as the beam intensity of accelerators increases, targets and beam intercepting devices will face several challenges such as high-energy densities (kJ/cm$^3$/pulse), high power densities (MW/cm$^3$), high averaged deposited power (MW), movable parts (rotating targets, for instance), physics requirements (materials with poor structural properties), radiation damage, and shorter lifetimes. Light source facilities face similar problems: as electron machines push towards higher brightness, the electron beam will have higher energy and power densities. Thus, collimators and beam dumps will face similar issues associated with high power targets.

The next generation of high power targets will use more complex geometries, novel materials, and new concepts (like flowing granular materials); however, the current numerical approaches will not be sufficient to converge towards a reliable target design that satisfies the physical requirements. We will discuss what can be improved in the next 10 years in target modeling to support high power (MW class) targets. This discussion is a result of the Letter of Interest campaign and workshops organized for SnowMass21.

## 2. Improving Codes Interfaces

Modeling high power targets is complex and requires a variety of skill sets. A target's three-dimensional geometry is first generated using CAD software. Then beam-matter interaction codes (e.g., MCNP, FLUKA, PHITS, Geant4) are used to simulate the interaction between the beam and the target matter. More specifically, the codes will predict the target production (e.g., muon, neutron, hadron), the energy deposition in the target, nuclide inventories after irradiation, and an estimation of the radiation damage in the material (displacement per atom or DPA). Finally, the heat depositions from these simulations are used with finite element analysis (FEA) and computational fluid dynamics (CFD) simulations to predict the thermo-mechanical response of the target (like in [4]). If the beam is pulsed, transient simulations are required and tend to be computing intensive. The accuracies of the FEA/CFD simulations are directly correlated with the particle simulation's accuracy, the material properties used (e.g., thermal conductivities, densities, sound speed), and the mesh used to model the geometry.

Consequently, the first challenge in target simulation is its interdisciplinary nature, which requires data exchange between different codes. For instance, particle and accelerator simulation codes are not commercial codes and lack common input/output standards. Thus, interfacing with them is generally laborious and limits considerably any optimization process. Most particle codes rely on their own geometry description and are not able to use the CAD model directly. Many efforts have been made to address this issue ([4], [5]); however, direct CAD imports are often unsuccessful and require the CAD geometry to be "repaired" manually so that it is watertight to be compatible with particle codes [4]. Another challenge is the data exchange between the beam-matter interaction codes and the finite element codes (FEA or CFD). Generally, a different mesh is used between the beam-matter interactions codes and the finite element codes. Thus, the definition of suitable mapping is critical for the accuracy of the FEA/CFD simulations. Similar efforts are being pursued in light sources to simulate the high-energy-density electron beams interacting with accelerator components such as collimators and beam dumps. To accurately model effects of losing a whole e-beam stored charge on the collimators as well as the resulting shower on downstream components, work is underway to couple the beam dynamics code Elegant with the particle-matter interaction program MARS and the MHD hydrodynamics code FLASH [6].

Due to the interfacing between the different codes, only design exploration is possible where a limited number of geometries are investigated. However, such an approach will be inefficient at higher power and may lead to targets not fulfilling the physics and mechanical requirements. Interfaces between the codes must be improved to allow a full parametric design optimization to develop robust design targets. Such an optimization attempt has been done recently [7] on the structural analysis of ORNL's SNS target and showed promising results. This type of optimization will also allow probabilistic design approaches that consider uncertainties such as beam profile, alignment, or geometric variation from manufacturing and would ultimately lead to robust designs [8].

## 3. Improving Simulations Accuracy

The simulations are useful only if they can accurately predict the physics of interest. The following discussion focuses on the inaccuracies related to the physics or approach rather than the solver algorithm's accuracy. As discussed earlier, a source of inaccuracy can come from the communication from one code to another. A slight error in the data mapping can lead to serious errors, especially with thin-walled targets. Thus, special care must be taken to ensure that any converter is accurate and allows proper data exchange. Another source of inaccuracy results when the code does not capture all the physics as is currently the case with the liquid metal targets, such as those with mercury. The presence of the inert bubbles in the liquid is not captured in the current simulations. Efforts are being made to capture more of the physics by proposing a constitutive model that can account for the volume taken by the oscillating bubble [9]. Still, to model the complex physics accurately, a coupled structure and fluid simulation is needed to capture the complex interaction between the compressible multiphase flow and the deformable steel vessel. Such Fluid Surface Interaction simulations are considered a grand challenge, especially since no good equation of state for mercury is available. Some physics are very complex to model, such as erosion and corrosion modeling due to cavitation or radiolysis. Another set of parameters that is rarely captured in current target simulations is the change of material properties in time due to radiation damage. For instance, it has been shown that neutron irradiation considerably changes the thermal conductivity of several materials such as graphite, ceramic, and copper [10] [11] [12]. Other mechanical degradation associated with irradiation that needs modeling includes radiation hardening, embrittlement, and swelling [13]. In order to estimate the target life span accurately, coupled transient simulation must be performed to evaluate the target thermo-mechanical response as its mechanical properties degrade with irradiation. Irradiation also makes fatigue modeling difficult. Although some data are available in the literature, more experimental data are needed for materials relevant to targets such as tungsten, tantalum, and other materials. Accordingly, a better understanding of radiation damage mechanisms is needed.

## 4. A Better Understanding of Radiation Damage

Irradiation of materials caused by particles produces atomic displacement damage that causes microstructural defects and nuclear transmutation that alter the material composition. Both the defects and the transmutation can then affect the material dimensions due to swelling; structural stability; mechanical properties (radiation hardening and embrittlement as mentioned earlier); corrosion resistance; and thermal, electrical, and optical properties. The radiation damage is estimated using particle codes with variables such as DPA in the molecular lattice structure which cannot be measured. Most of our radiation knowledge comes from post-irradiation examination (PIE) experiments in which a sample is irradiated and then characterized by microscopy and thermo-physical and -mechanical tests. Although this approach is the gold standard, it has its limitations. When the sample gets irradiated with protons or neutrons, it becomes radioactive and a Hot Cell is required to manipulate the sample, or a long period of time is needed to let the sample cool off. Because only a few facilities in the world provide such services, testing a large number of samples becomes prohibitive very quickly. Thus, there is a compelling need to develop numerical tools that can emulate the radiation damage from the atomic to the macroscopic scale. Quantum mechanical or molecular dynamics simulations can be used to understand the physics at the atomic or nanoscale level, but they quickly become too computationally demanding at the needed length and time scale (cm, days). Thus, phase-field and continuum models that leverage these small-scale simulations must be developed to simulate radiation damage on the macroscale. For instance, a swelling model has been developed based on cluster dynamics simulations of defect evolution in materials [14]. More efforts are needed in that direction, with experiments [16] in parallel to validate the multiscale approach. The RaDIATE collaboration has been performing such work with the University of Wisconsin and Pacific Northwest National Laboratory, but further work is needed to predict the physio-mechanical properties of the irradiated material. Such a numerical tool would also allow the exploration of novel materials for targets that currently cannot be considered because of the lack of irradiated materials data.

5. **Better Verification and Validation**

Although target design relies heavily on simulations, the targets are generally insufficiently instrumented to perform a satisfying verification and validation (V&V) of the numerical approach. Sensors that can survive such radiation are limited, and disposing of these sensors at the end of the target life can be challenging. Because of the lack of instrumentation, conservative approaches (assumption of higher heat deposition or overfocused/defocused beam profiles) are used to consider the simulation uncertainties. As the beam energy increases, these conservative approaches may not be capable of leading to a design. A better validation of the numerical approach must be performed in order to gain more confidence in the numerical results. Fully instrumented test targets with simpler geometries need to be used to validate the numerical approach. Facilities such as Los Alamos Neutron Science Center (LANSCE) or the High-Radiation to Materials (HiRadMat) Facility at CERN can be used to test these targets, and the Target System Integration Building (TSIB) High Power Targetry (HPT) Lab facility, under development at Fermilab, could be used for PIE to validate the radiation damage model used. Beam dumps can also be used for such validation: the lack of moving parts and generally simple geometry makes them excellent candidates for such an effort.

6. **Leverage Advanced Concepts and Future Computing**

The upcoming exascale supercomputers (such as Frontier and Aurora) will start a new era in modeling, delivering a computing power of $10^{18}$ floating point operations per second by leveraging extreme parallelism from heterogeneous central processing unit (CPU)/graphics processing unit (GPU) architecture. To effectively use this computing power, new algorithms and software have been developed in a broad range of applications such as biochemistry, computational fluid dynamics (NEK5000), molecular dynamics (GROMACS), and neutronics. The target community will need to work closely with the computing community to understand how to leverage these computing tools for better target design. Such computing power could facilitate predicting the target performance over time, as its material properties are degrading due to radiation damage. It could also support performing uncertainty quantification simulations to better understand the high cycle fatigue on some of the targets to predict their life span more accurately.

Over the last few years, a new trend in computing has been to leverage reduced-order modeling (ROM) to build digital twins of engineering designs. This digital twin approach is becoming increasingly relevant to model-based system engineering such as targets. A digital twin is a virtual prototype of a physical system that is continually updated with measurements from the physical system (the physical twin). In particular, the digital twin of a target could help determine the life span of the target, understand the target's performance, refine the modeling assumption (like beam profile), reflect the target's age (material properties that degrade with radiation damage), and potentially detect imminent failure. Target and other engineering systems (e.g., cryogenic process loop, water loop, dumps) could leverage the models that can be used to validate the system model with measurements, provide decision support, and predict changes in the physical system over time.

Simulating the billions or more particles in a beam or the billions of molecules in a material will remain challenging even using the state-of-the-art exascale supercomputer. Quantum computing could be a great opportunity to determine molecular and particle codes. Quantum computers use the properties of quantum states to perform calculations. Unlike classical computers, which use binary bits (0 or 1 states only), quantum computers use qubits that can be in a 0 or 1 quantum state or a superposition of both. In particular, quantum computers are believed to be capable of solving certain problems that would take years on classical computers. Although quantum computers will not replace classic computers in the next decade, the quantum and high energy physics communities should foster their collaboration to determine how quantum computing could address scientific challenges in high energy physics.

7. **Conclusion**

Simulations play a critical role in target design, and as beam power increases, better numerical tools will be needed to keep the target reliable and cost-effective. The target community has been collaborating through the High Power Targetry Workshop since 2003, discussing the operation and simulation challenges with the targets. Such international collaboration will be essential over the next 10 years to expand our knowledge associated with high power targets. Accelerator facilities and GARD (DOE HEP General Accelerator R&D) will need to invest more in target R&D to better understand radiation damage and bring target modeling to the needed level for high power accelerators.